# Cloud Broker: A Systematic Mapping Study

Hoda Taheri
Computer Engineering Department, Ferdowsi University of Mashhad, Mashhad, Iran, h.taheri@mail.um.ac.ir

Faeze Ramezani
Computer Engineering Department, Ferdowsi University of Mashhad, Mashhad, Iran, fa.ramezani@mail.um.ac.ir

Neda Mohammadi
Computer Engineering Department, Ferdowsi University of Mashhad, Mashhad, Iran, n.mohammadi@mail.um.ac.ir

Parisa Khoshdel
Computer Engineering Department, Ferdowsi University of Mashhad, Mashhad, Iran, p.khoshdel@mail.um.ac.ir

Bahareh Taghavi
Computer Engineering Department, Ferdowsi University of Mashhad, Mashhad, Iran, Bahareh.taghavi@mail.um.ac.ir

Neda Khorasani
Computer Engineering Department, Ferdowsi University of Mashhad, Mashhad, Iran, neda.khorasani@mail.um.ac.ir

Saeid Abrishami (corresponding author)
Computer Engineering Department, Ferdowsi University of Mashhad, Mashhad, Iran, s-abrishami@um.ac.ir

Abbas Rasoolzadegan
Computer Engineering Department, Ferdowsi University of Mashhad, Mashhad, Iran, rasoolzadegan@um.ac.ir

**Abstract**—In a cloud environment, a cloud broker is an important entity that works as an independent middleware between cloud customers and providers to address issues and conduct negotiations related to satisfying both customer preferences and service provider profits. In recent years, researchers have published many articles which directly or indirectly address this research area. A systematic method is vital for extracting all search spaces (journals, conferences, and workshops) and primary studies (articles) conducted in the cloud broker field and then selecting some of the highest quality studies. An important part of a systematic review is its provision of an appropriate research method that can extract large volumes of related studies. The proposed systematic review includes a comprehensive three-tier search strategy (manual search, backward snowballing, and database search). The accuracy of the search methodology has been analyzed in terms of extracting related studies and collecting comprehensive and complete information in a supplementary file and also, the detailed explanation of the reviewing process is inserted in Appendix A. In the search methodology, qualitative criteria have been defined to select studies with the highest quality and the most relevance among all search spaces. Furthermore, to find studies related to the cloud broker field, some queries have been created that employ relevant keywords in that field. In the present study, out of 1,928 extracted search spaces, 171 search spaces have been selected based on the defined quality criteria. Then, 1,298 articles have been extracted from these 171 selected search spaces. As a result, 496 high-quality papers have been selected among the mentioned papers. The chosen papers were published in prestigious journals, conferences, and workshops from 2009 through 2019. In the current Systematic Mapping Study (SMS), eight research questions have been designed for the purpose of identifying information that is significant to the cloud broker field, such as the most critical and debated topics, existing trends and issues, active researchers and countries, commonly used techniques in building cloud brokers, evaluation methods, the amount of research conducted by year and the place of publication, and the most important active search spaces. This information has been extracted from 496 selected papers (their references are in Appendix B) and can provide a useful guide for research teams and developers interested in this field.

# 1 INTRODUCTION

In the cloud environment, cloud services comply with pay-as-you-go logic, meaning that each cloud customer should pay for as much as it consumes [1]. Cloud services offer several benefits, such as high availability, flexible application deployment, and low cost. Nevertheless, the role of cloud brokers is still in the infancy stage [2], [3]. The market of cloud services consists of a huge number of services, many of which share the same functionality but with different degrees of quality. Therefore, cloud customers face a huge challenge when selecting proper services according to their preferences. A broker can play a vital role in overcoming this challenge by negotiating between all providers and cloud customers to find the most suitable services that consider customer preferences and provider profits [3],[4]. Overall , cloud broker is an entity that manages the use, performance, and delivery of cloud services and negotiates relationships between cloud providers and cloud consumers [NISTIR 8006 from NIST SP 500-292].

In some situations, cloud customers may become dependent on a particular cloud service provider. This is known as the vendor lock-in problem, in which customers cannot easily move between cloud providers without paying an extra cost. A broker can help customers avoid vendor lock-ins. Such benefits lower the cost of offering services for providers and create a flawless switch between cloud providers that satisfies customer preferences [5]. To actualize the function of the broker, applications should be able to remove the limitations of each cloud provider and thus provide cross-cloud computing [3] aimed at supporting developers with challenges related to interoperability, migration, resource planning strategies, and dynamic deployment. Usually, the term "broker" has been utilized to depict various intermediation models. One of these models is a cloud federation that creates a common technology for implementing cloud services which all providers must obey. In contrast, a multi-cloud model does not consider any common technology. Therefore, for customers to switch among cloud providers, the broker must first solve the differences among all providers [2]. Due to the crucial role of the broker, the past decade has seen a large volume of research focused on investigating different broker responsibilities.

**NIST has classified services offered by cloud brokers into three categories:** arbitration, aggregation, and intermediation [6]. In cloud **service aggregation,** multiple cloud services have been combined and aggregated into one service. The broker is responsible for providing data security when data is transferred between the cloud customer and multiple cloud providers [6]. In aggregation, two services or more have been aggregated into a single service to increase broker capabilities [7]. Because it allows services to be selected from among different providers, cloud service arbitration is more flexible than service aggregation. In other words, in **service arbitration**, the broker can select services from various providers based on the data's characteristics or the context of the service [6]. In cloud **service intermediation**, a cloud broker enhances a given service by improving some specific capability and providing

- *The authors Hoda Taheri, Faeze Ramezani, Neda Mohammadi, Parisa Khoshdel, Bahareh Taghavi, Neda Khorasani, Saeid Abrishami (∗corresponding author), and Abbas Rasoolzadegan are with the Computer Engineering Department, Ferdowsi University of Mashhad, Mashhad, Iran.*
- *E-mail addresses: {h.taheri, fa.ramezani, n.mohammadi, p.khoshdel, Bahareh.taghavi, neda.khorasani}@mail.um.ac.ir and {s-abrishami, rasoolzadegan}@um.ac.ir*

value-added services to cloud consumers [6]. A comprehensive and systematic review of the research in this field is crucial to identifying its major trends and issues. Such a review can act as a guideline for all researchers and enthusiasts in their search for a deeper understanding of the challenges and issues needing to be addressed. An accurate methodology is critical for covering and reviewing all relevant high quality research studies. This methodology should offer some important features, such as reliability, impartiality, in addition to the traceability of results. However, there are few studies that systematically investigate and analyze the area of cloud brokerage. In 2005, Deba et al. [8] introduced an evidence-based software engineering method that consists of two well-known methodologies, i.e., the Systematic Literature Review (SLR) and the Systematic Mapping Study (SMS) [9],[10]. For searching through all research works and reviewing them, both SLR and SMS feature the same methodology [11],[9],[10]. However, Zhang and Budgen [9],[10],[12] illustrated some differences between these two methodologies. The major difference is in the determination of the final goal. Indeed, both methodologies pose different research questions that should be answered at the end of the review. It can be stated that the SMS research questions are more general and work toward the goal of identifying research trends and topics in the specified field. In contrast, SLR tries to extract the data from the initial studies and subsequently answers some specified RQs (Research Questions) [9].

In order to have proper coverage of the research studies conducted in the cloud broker field, a systematic methodology is essential for detecting all search spaces (i.e., journals, conferences, and workshops) and research studies in this area. Hitherto, we have conducted a SLR and SMS on different research fields [13],[14],[15]. In 2017, there was a search methodology to review studies on software design patterns [14] and then, in following works [13],[15], the methodology was partially improved. Ahmadian et al. used a comprehensive search methodology to review the intrusion alert analysis in intrusion detection systems [15]. In another review study, Javan et al. systematically covered research studies on security patterns in software design [13]. As previously mentioned, a primary part of this methodology is presenting an appropriate search strategy to extract related research studies in the field under study. The proposed systematic review is an extended version of the current authors' previous works [13],[15]. **A three-tier search strategy is used in the present SMS which comprises a manual search, backward snowballing, and a database search.** In the **manual search phase**, each venue (journals, conferences, and workshops) in the search space list (acquired from investigating the references of existing reviews in Table 1) is manually searched using a set of constructed queries. Each query consists of a set of keywords seen in Table 2 illustrates the constructed queries for finding related papers and studies. The goal of the **backward snowballing phase** is discovering some new papers which were not found in the previous phase. In the backward snowballing phase, the references of all currently included papers are investigated. It should be noted that, in each phase of the search for new studies, papers are evaluated in terms of quality and only the sets of papers with the desired quality are selected and placed in the set of included papers. In Section 2, the criteria for evaluating all new search spaces and new papers are explained in detail. In the **database search phase**, a manual search is conducted by employing defined keywords in $SuppFile_{W3,T3}$ on well-known databases, such as Google Scholar, Springer Link, IEEEXplore, ACM Library, and ScienceDirect. By applying all search phases together, a large dataset is extracted composed of 1,298 papers related to the cloud broker field and published from 2009 through 2019. Following this, 496 papers are selected (the references are in Appendix B) according to some inclusion and exclusion criteria and, finally, the selected papers are analyzed and used to answer eight research questions. The following sections provide the details of this process.

In the presented search methodology, some qualitative criteria have been defined for selecting the highest quality and most relevant articles among all search spaces and studies. In addition, to find studies related to the cloud broker field, eight queries have been designed based on keywords in the field under study. The accuracy of the search methodology in finding related studies has been computed. For more clarity, a supplementary file, entitled *SuppFile*, has also been created which consists of document files providing information about search spaces, extracted papers, keywords, and RQs. Also included in these document files is a description on how the process of the three-tier search strategy (manual search, backward snowballing, and database search) has been carried out by the research team. The *SuppFile* has all of the gathered data in the SMS and so holds comprehensive knowledge about all data in the desired field. A quick guideline to utilizing *SuppFile* is available in readme.pdf file. The present paper may refer to some data, tables, and other information from *SuppFile*. For example, $SuppFile_{E1,T1}$ refers to Table 1 of Excel file $E1$, or $SuppFile_{W1,T5}$ refers to Table 5 of document $W1$ in the *SuppFile* folder. This file is available along with two appendixes A and B at the address https://github.com/hodataheri/SMS-CloudBroker.

In this Systematic Mapping Study (SMS), eight research questions have been designed to identify information relevant to the cloud broker field, such as the most critical topics, current trends and issues, active researchers and countries, commonly used techniques in building cloud brokers, evaluation methods, the amount of research conducted by year and the place of publication, and the most important active search spaces. This information can provide a useful guide for research teams and developers interested in the desired field. An SMS can be used as a pre-review before conducting an SLR to gather general information on the desired research topic. As seen in Table 1, for the starting point of an SMS on the cloud broker, 24 secondary studies (survey and review) are randomly selected to investigate the literature relevant to cloud brokers. It should be noted that, during the research methodology, other review studies are found which are seen in $SuppFile_{E2,T1}$.

This paper is organized as follows. Section 2 presents the research methodology used in this SMS. Section 3

provides the results of the current systematic study that have been categorized by eight research questions. In Section 4, the acquired results of Section 3 are discussed. Based on the findings of the current SMS, Section 5 presents some implications for researchers, stakeholders, practitioners, and educators interested in the field under study. Finally, Section 6 provides the conclusion.

## 2 RESEARCH METHODOLOGY

Distinct works already exist on conducting an SMS and designing a unique research methodology [16],[17]. One of the most prominent of these research methodologies is by Peterson et al. [16]. Along with some updates and improvements in some phases, the current study's research methodology is adapted from the three SMSs performed by Peterson et al. [16], Ramaki et al. [15], and Javan et al. [13]. Considering the investigation of research studies, it is worth mentioning that the advent of the cloud broker was in 2009. Therefore, the present research's SMS covers all research articles published from 2009 through 2019, out of which works relevant to the cloud broker field undergo a complete review. It is important to note that the field of security can be considered separately as it is a very broad area. For this reason, the proposed SMS does not address articles related to security and such articles are excluded during the search process for the selection of related studies. Generally, the SMS introduced by the current study consists of two main steps: the planning step and the conducting step. The following briefly describes each of these steps while Appendix A provides a complete explanation.

### 2.1 The planning step

The planning step comprises seven fundamental phases. The **first phase** specifies the scope and research questions. Defining the research questions determines the research goals, which are met while the SMS is conducted. In the proposed SMS, eight comprehensive and different RQs are defined. Responding to these can cover all of the objectives. Table 3 describes these RQs and explains the rationality for each. The **second phase** specifies the search strategy. In this phase, three search strategies (i.e., manual search, backward snowballing and the database search) are applied to find the studies related to the cloud broker field.

The **third phase** specifies the search space. In fact, search spaces are publishers who have published relevant studies. At the beginning of the review process, the search space set is empty. Therefore, as described in Section 1, a set of secondary studies (i.e., surveys and reviews), shown in Table 1, are randomly selected to begin the review process. The list of search spaces is quantified by reviewing the cited articles in the reference section of the secondary studies. The **fourth phase** specifies the search string. After the search space list is quantified, it is necessary to begin searching in the search spaces for studies related to the cloud broker field through the use of search strings. Therefore, at this stage, related keywords are merged and queries are created. These queries, which are the same as search strings, are used to search for related studies in the search spaces. Table 2 presents the queries employed in the proposed SMS.

The **fifth phase** plans the study selection process which leads to determining the studies to be included. Out of the studies found during the previous phases that are related to the cloud broker field, those having the necessary qualifications are reviewed and analyzed. For this purpose, a set of quality criteria is defined, according to which studies published in journals, conferences and workshops are qualitatively reviewed and selected. Appendix A provides details and tables related to these quality criteria. Tables 4, 5, and 6 illustrate the exclusion criteria for journals, the exclusion criteria for conferences and workshops, and the exclusion criteria for the extracted studies, respectively. Appendix A presents a full description of these exclusion criteria

Table 1. Secondary studies for generating the initial sms set

| No. | Secondary Study Title | Year | Ref. |
|---|---|---|---|
| 1 | Brokering in Interconnected Cloud Computing Environments: A Survey | 2018 | [7] |
| 2 | A Review on Service Broker Algorithm in Cloud Computing | 2017 | [18] |
| 3 | A Comprehensive Study on Cloud Service Brokering Architecture | 2017 | [19] |
| 4 | Cloud Services Recommendation Reviewing the Recent Advances and Suggesting the Future Research Direction | 2017 | [20] |
| 5 | Service Provisioning in Cloud: A Systematic Survey | 2017 | [21] |
| 6 | A Survey on Various Cloud Aspects | 2016 | [22] |
| 7 | A Classification and Comparison Framework for Cloud Service Brokerage Architectures | 2016 | [16] |
| 8 | A Review on Broker Based Cloud Service Model | 2016 | [17] |
| 9 | Cloud Service Brokerage: A Systematic Literature Review using a Software Development Lifecycle | 2016 | [23] |
| 10 | Resource Provision Algorithms in Cloud Computing: A Survey | 2016 | [24] |
| 11 | Towards a Holistic Multi-Cloud Brokerage System: Taxonomy, Survey and Future Directions | 2015 | [25] |
| 12 | A survey on SLA-based Brokering for Inter-Cloud Computing | 2015 | [26] |
| 13 | Cloud Services Brokerage: A Survey and Research Roadmap | 2015 | [27] |
| 14 | Cloud Service Selection: State-of-the-art on Future Research Directions | 2014 | [22] |
| 15 | Cloud Computing Service Composition: A Systematic Literature Review | 2014 | [28] |
| 16 | A Comparative Study of Traditional Cloud Service Providers and Cloud Service Brokers | 2014 | [29] |
| 17 | A Review of Literature on Cloud Brokerage Services | 2014 | [30] |
| 18 | A Literature Review on Cloud Computing Adoption Issues in Enterprises | 2014 | [31] |
| 19 | A Survey on Needs and Issues of Cloud Broker for Cloud Environment | 2014 | [32] |
| 20 | Survey on important Cloud Service Provider attributes using SMI Framework | 2013 | [33] |
| 21 | A Comparison Framework and Review of Service Brokerage Solutions for Cloud Architectures | 2013 | [34] |
| 22 | A Survey on Interoperability in the Cloud Computing Environments | 2013 | [35] |
| 23 | A Survey on Infrastructure Platform Issues in Cloud Computing | 2012 | [36] |
| 24 | Inter-Cloud Architectures and Application Brokering: Taxonomy and Survey | 2012 | [37] |

Table 2. Constructed queries for finding related studies

| # | Query |
|---|---|
| 1 | Cloud broker |
| 2 | Cloud AND service (arbitration OR intermediation OR aggregation OR integration OR customization OR Orchestrat) |
| 3 | (Multi cloud OR Federat cloud OR Cross cloud OR Inter cloud OR (third party AND Cloud))AND (auction OR negotiation OR pricing OR interoperability OR management) |
| 4 | Cloud AND ("service composition" OR "service selection") |

Table 3. Defined research questions

| # | Research Questions | Rationale |
|---|---|---|
| 1 | How active is the field of brokering and how is the distribution of selected studies by type over publication year (journal, conference, and workshop)? | Detecting the current volume of research and primary trends better discern the attraction of the field. This requires investigating the annual publication volume of research studies in the field. |
| 2 | Which researchers and research venues are more active in this field and how are the active researchers distributed geographically? | The demographics of brokering research techniques provide a useful starting point for interested researchers by identifying active scholars, venues, and countries. |
| 3 | What are the core research topics in the field of brokering? | Identifying and classifying the current research in terms of brokering techniques. Analyzing the evolution and distribution of each topic and the potential trends in the research focus. |
| 4 | Which broker topics have the least/most corresponding attention and what is the publication trend and distribution for each topic? | Some objectives might be more prominent than others, but broker developers should take care to cover a varied spectrum of topics. |
| 5 | Which forms of empirical evaluation have been used? What are the tools available to support field approaches? Which techniques are used more in the field? | Empirical evaluation indicates whether the environment is real or if simulation and supporting tools can describe frameworks, platforms, or simulation. Techniques can be game theory, optimization, and heuristic. |
| 6 | What is the relation between topics and broker roles in the NIST category? Which NIST roles have the least/most corresponding attention? | General classification schemes might work to an extent, but a precise and comprehensive classification of broker roles should address broker-specific criteria. |
| 7 | In which environment and service layer is the broker mostly considered? | Environments are multi-cloud, federated, etc. and the service layer is IaaS, SaaS, PaaS, and XaaS. |
| 8 | What is the broker's control orientation? | Generally, types of control orientation are centralized and distributed. |

Table 4. Journal exclusion criteria (JEC)

| # JEC | Description |
|---|---|
| 1 | If the journal is not indexed in the JCR |
| 2 | If the scope of the journal is not related to our desired field |

Table 5. Other exclusion criteria (OEC)

| # OEC | Description |
|---|---|
| 1 | ((Qualis<A5) OR (ERA<A) OR [(H5_Index<15) AND ((Qualis<A5) OR (ERA<A))] OR (Metrics Not Available)) |
| 2 | (Aims and scopes are not related) |

Table 6. Exclusion criteria for extracted studies

| # | Description |
|---|---|
| 1 | The Study is not a primary study (e.g., survey) |
| 2 | Study cannot be accessed (e.g., book chapter) |
| 3 | The Study is out of our primary scope (e.g., security aspects in broker) |
| 4 | The Study belongs to an excluded search space (according to Tables 4 and 5) |
| 5 | The Study's contribution is not related to the cloud broker (i.e., the broker just mentioned in the study but it is not the goal) |

A complete list of all extracted search spaces and reasons for the exclusion and inclusion of journals and others (conferences, and workshops) is provided in $SuppFile_{W3,T4}$ and $SSuppFile_{W3,T6}$, respectively. Moreover, the aims and scope of each of the search spaces are introduced in $SuppFile_{W3,T5}$ and $SuppFile_{W3,T7}$.

$SuppFile_{E10,T3}$ and $SuppFile_{E10,T6}$ present a list of extracted papers and comprehensive information about the reason for the exclusion of each extracted paper. The main aims and scope of the cloud broker field are specified in $SuppFile_{W3,T1}$.

After the studies to include are determined (the references of 496 included studies are in Appendix B), the keywords of the included papers are investigated and, as a result, some new keywords are extracted and added to the set of keywords for later use. $SuppFile_{E10,T3}$ and $SuppFile_{E10,T6}$ provide the complete information of all other papers (conferences and workshops) and journal papers respectively. $SuppFile_{E1,T5}$ presents the entire list of all included papers (journals, conferences, and workshops). Some valuable points are explained in the following.

- The field of the cloud broker comprises a large pool of research studies. Therefore, some thresholds are empirically selected for the exclusion criteria which consider two principals: 1) a small change in the exclusion thresholds should not significantly affect the number of excluded/included papers and 2) the application of these thresholds should not exclude a large number of highly cited papers.
- Although the desired field in the proposed SMS is the cloud broker, search spaces with the scope and aim of web services or distributed and parallel computing are also considered. Also included are search spaces in other research areas, because they have published papers in the field of cloud computing. An example is IEEE Transactions on Smart Grid.
- There are some search spaces, such as the International Conference on Software Engineering (ICSE), that have scarcely published any papers in the field of cloud computing as their scope is software engineering. Such search spaces are, therefore, excluded.
- In rare cases, the present study found a search space with two names. For instance, the ACM International Symposium on High-Performance Distributed Computing and the International Symposium on High-Performance Parallel and Distributed Computing are two names for the HPDC search space. Also found was a search space whose name has changed. The initial name for GLOBECOM search spaces was IEEE Global Telecommunications Conference, but, after 1972, its name was changed to the Global Communications Conference. Another example are two search spaces, namely the European Conference on Machine Learning (ECML) and the European Conference on Principles and Practice of Knowledge Discovery in Databases (PKDD), which, after 2007, merged and were entitled the European Conference on Principles and Practice of Knowledge Discovery in Databases (PKDD).
- It should be noted that each venue appears just once in the search spaces, meaning that annual workshops or different editions of conferences or journals issues are not repeated.

The **sixth phase** specifies the search and study evaluation strategy. In this phase, the goal is to examine the completeness of the search strategies utilized for finding the related studies. Therefore, objective evaluations (i.e., quantitative criteria) and subjective evaluations (i.e., managed by the expert(s)) are employed for this evaluation. The proposed search strategy review utilizes these two prominent and most valid metrics for this purpose. Eq. (1) and Eq. (2) perform the objective evaluations and subjective evaluation respectively. To achieve a more objective evaluation, the quasi-sensitivity metric is selected to evaluate the applied search and study selection.

$$Sensitivity = \frac{Number\ of\ studies\ in\ the\ proposed\ SMS}{Number\ of\ studies\ overall} * 100 \quad (1)$$

$$QGS = \frac{The\ number\ of\ discovered\ papers\ in\ the\ search\ strategy}{The\ number\ of\ discovered\ papers\ in\ the\ evaluation\ phase} * 100 \quad (2)$$

where the strategy search is the three-tier process, namely the manual search, backward snowballing search, and database search.

In the evaluation phase, the home pages of active researchers in the area of cloud computing are visited and any of their papers relevant to the cloud broker field are extracted. After the extraction of these studies from the home pages and the application of the inclusion/exclusion criteria, the remaining unseen papers comprise the present study's Quantitative Gold Standard (QGS), according to Eq. 2. The aim of applying the QGS is to calculate the quasi-sensitivity and then compare the obtained result with a predefined threshold. Accordingly, if the result falls below the threshold, the search and study selection process should be repeated using the QGS. By observing [20], an acceptable threshold should be between 70% and 80%. In the evaluation phase, 32 articles were found, of which 26 articles were previously identified in the main search phases of the systematic review process. Thus, a sensitivity of 81.25% was achieved, which is above the current research's predefined threshold. In other words, the probability of not finding a paper related to the cloud broker was less than 20%. It can then be concluded that the results acquired from the proposed review have satisfactory accuracy and validity. Section 1.6 of Appendix A presents the details of the steps taken to evaluate the search strategies and the metrics for calculating the accuracy and completeness of these strategies so as to extract the related studies.

The **seventh phase** plans the data extraction and classification process. Here, the data extraction forms are specified and also the data extraction strategy is determined. To prepare useful information to answer the RQs, the type of data extraction and strategy should be specified. After the determination of which studies are included, based on the defined quality criteria, different parts of the articles (i.e., the title, abstract, keywords, and body) are examined to extract the information needed for later analysis. The extracted data are organized into tables and utilized as the information for responding to the RQs of the SMS ($SuppFile_{E1,4\ to\ 10}$ includes all the information needed to answer the RQs).

## 2.2 The conducting step

After the planning step, if the mapping study reaches an acceptable level of quality in the evaluation phase, then the conducting step can commence. First, the secondary studies presented in Table 1 conduct the search and study selection phase. These articles are found by a manual search process that employs the general term of *cloud broker* along with keywords, such as *review* or *survey*. Then, the cited papers of the secondary studies are extracted as the present study's initial set. The three-tier search strategy is conducted here to complete the result set. As previously mentioned, a sensitivity of 81.25% is achieved, which is above the current work's predefined threshold. The data extraction phase is then conducted. After the determination of which studies are to be included, based on the defined quality criteria, information is extracted from this collection of included papers to answer the research questions. This information is stored in the tables and, to extract such information, different parts of the article, such as its title, abstract, keywords, and body, are examined. After being organized into tables, the data should be accurately investigated and analyzed in order to respond to the RQs of the SMS. The main topics of the cloud broker's field are determined by the extracted keywords of the included paper. The aggregated keywords are classified semantically through 18 steps which create the research tree shown in Fig. 1.

The research tree is a multilevel structure determining the most important topics in the desired field. The primary topics are located at the top of the research tree and subtopics are located in other levels, even as leaves. The location of a topic in the research tree is dependent on the number of its repetition in the included studies. Sections 3.4 in Appendix A provides a complete explanation for finding important topics and determining the steps for building the research tree.

The examination of related topics by experts and their aggregation finally creates 10 main topics in the cloud broker field that are introduced as level-one topics of the research tree. The topics are divided into two categories: client-centric and provider-centric. Client-centric topics are activities that the cloud broker performs in response to a user request. For example, due to a user's service request, the broker may perform service discovery, service selection, service composition, and so on. Provider-centric topics are activities performed by the cloud broker based on the provider's request, for example, pricing, resource allocation, or energy management. It should be noted that service allocation includes service provisioning and scheduling on the client-side. Fig. 1 depicts the research tree created in the present SMS. The percentage of included papers in each topic is shown below the topic. One of the rubrics applied for the evaluation of the proposed SMS addresses threats to its validity. In the validation process, the primary goal is to provide some evidence for resolving all existing threats to the proposed SMS. The following investigates and discusses the main evidence.

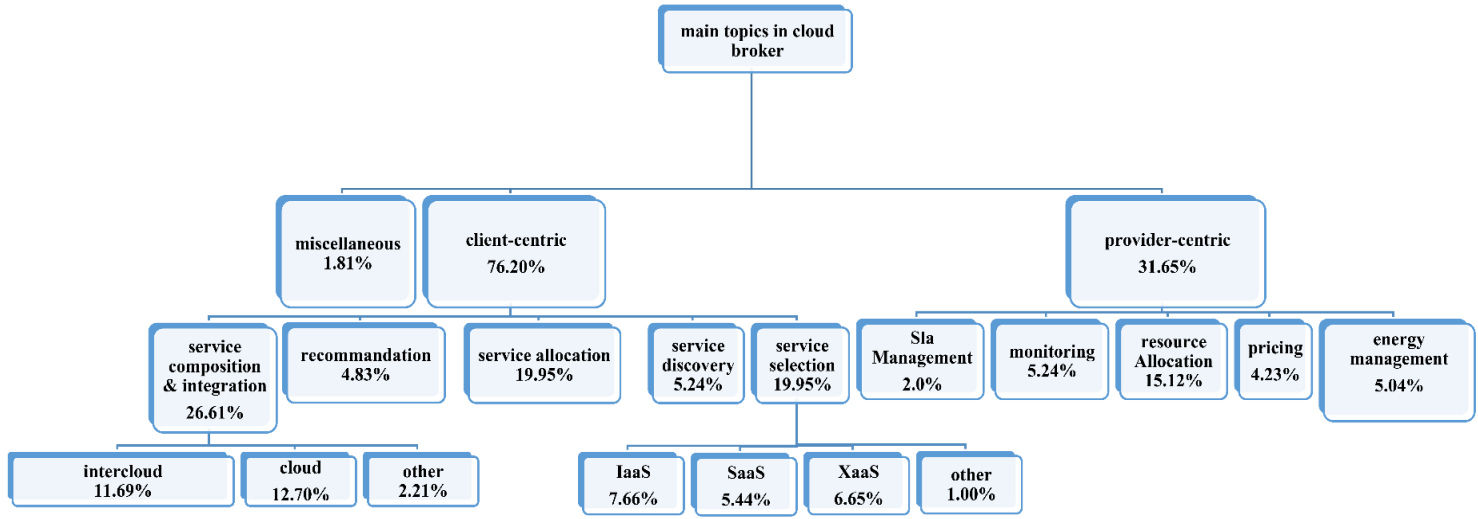

Fig. 1. Multilevel research tree containing the main topics and sub-topics in the field

- **Obtaining a set of high-quality studies**: For obtaining a complete set of high-quality studies in the cloud broker field, a complete procedure has been designed as the proposed search strategy. This offers the advantages of the two famous search methods, i.e., SLR and SMS. For this reason, the present review is believed to be reliable.
- **Obtaining the most related studies**: One of the most primary advantages in the current study's search strategy is its gradual evolution of a keyword set while the SMS is conducted. It should be stated that, in some situations, some keywords in the set of keywords do not convey the concept of the cloud broker. Accordingly, all such keywords in their category are merged by applying some logical operators (AND, OR). $SuppFile_{E3,T2}$ provides some examples of merging keywords.
- **Reviewer's biases or misunderstandings during the process of study selection**: To prevent these biases or misunderstandings challenges in the proposed SMS, the selection process is first independently performed by two reviewers. Then, any possible disagreements are solved by the third reviewer and the resulting decision rules.
- **Creating some forms to extract raw-data**: During the execution of the data extraction phase, another threat is posed when some of the included studies do not have any of the designated keywords. By considering the context of these studies, the current work extracts suitable keywords and stores these in the related forms.
- **Assigning proper name for each level-one topic**: After clustering all keywords (extracted and generated), the present study assigns a proper name for each cluster that describes that cluster's concept. For example, for the topic of Resource Allocation, there are some keywords, such as Resource Allocation, Resource Management, and VM Scheduling. However, Resource Allocation is selected as the suitable name for the topic, because it best conveys the desired concept of the studies. Nevertheless, in addition to considering both the semantics and syntax of the keywords for each topic, the proposed SMS also introduces the selected names to its reviewers to remove any potential bias.

## 3 RESULTS OF THE STUDY

The previous sections explain in detail the process of searching for study and space. This section shall analyze and respond to the RQs presented in Section 2 according to their arrangement in Table 3. It is worth noting that the level-one topics in the research tree cover all the sub-topics in the lower levels. For this reason, an analysis is carried out on these topics.

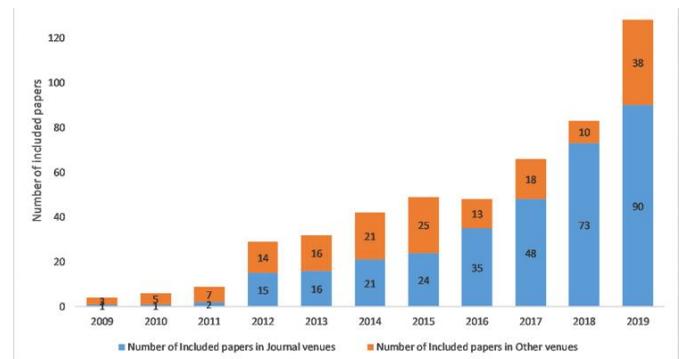

Fig. 2. Number of included papers per year

### 3.1 How active is the field of brokering and how is the distribution of selected studies by type and publication year (journal, conference, and workshop)? (RQ1)

One of the primary RQs calls for investigating the frequency of published papers in the field of cloud brokers from the advent of cloud brokers (2009) through the end of 2019. In consideration of the studies extracted by the queries defined in Table 2, it can be observed that the time interval of studies published in the field of cloud brokers is from 2009 to 2019. This observation points to the acceptance level and progress of the cloud broker research field during the period under review. The current study illustrates the level of attention of academic societies and its changes over these years. Through an analysis of the included study set, a logical trend in published research is recognized which highlights the responsibilities of brokers in the cloud environment. Fig. 2 illustrates the final results. The data depicted in Fig. 2 to 5 is acquired based on analyzing the information inserted into $SuppFile_{E1}$ and $SuppFile_{E4\ to\ E10}$. As shown in Fig. 2, among 496 included studies, 55% were published between 2017 and 2019. With the advent of the multi-cloud concept, the number of papers in this field has increased since 2012 and has grown further in recent years. In 2019, the broker field reported the highest number of studies compared to previous years. Fig.3 shows the frequency of top-level topics in the included studies between 2009 and 2019. As seen, some topics have a higher frequency in these articles. *Service Composition and Integration*, *Service Selection*, and *Service Allocation* have the highest frequency in the included studies. The reason why more attention is paid to the concepts of service composition and service selection is their complexity. In general, subjects that are scientifically more complex are more popular. Moreover, these topics are the main services of a cloud broker according to the NIST definition. In contrast, more technical and less complex subjects, such as service discovery and monitoring, have fewer papers. By the way, some topics, such as pricing and recommendation, have a high chance of being investigated in the future.

From another point of view, the importance of the topics extracted can be examined by considering the queries utilized in the search phase for related studies. Fig. 4 presents the importance of the queries defined in Table 2 in retrieving related papers and illustrates that the queries are well designed and make an equal contribution (approximately 25%) to the results.

It is worth noting that, of the 496 included studies, 23% (115 studies) were retrieved through the first query; of these, 21% (101 studies) were extracted directly in the first query and 2% of the remaining studies were extracted by a combination of the first query and other queries. The highest overlap was between Queries 1 and 3, with eight included studies. Another type of analysis of the included studies is based on search spaces. Among 496 extracted studies, 326 studies were published in journals and 170 in conferences and workshops. Table 7 and Table 8 list the most important journals and conferences respectively.

### 3.2 Which researchers and research venues are more active in this field and how are the active researchers distributed geographically? (RQ2)

Knowing the active researchers in the field of cloud brokers is useful for those researching and working in this area. The results acquired in the evaluation phase, found in $SuppFile_{E1}$, are used to extract the active researchers. $SuppFile_{E1,T2}$ provides the list of active researchers and the country of the authors are found in $SuppFile_{E1,T1}$ and $SuppFile_{E1,T4}$.

Knowing the active researchers in the field of cloud brokers is useful for those researching and working in this area. The results acquired in the evaluation phase, found in $SuppFile_{E1}$, are used to extract the active researchers. $SuppFile_{E1,T2}$ provides the list of active researchers and the country of the authors are found in $SuppFile_{E1,T1}$ and $SuppFile_{E1,T4}$. Fig. 6 illustrates the active researchers according to the number of publications in the field of cloud brokers. Researchers rank in descending order based on the number of publications from left to right. For example, with 22 publications in this field, Rajkumar Buyya is placed at the top of the list. It should be noted that the more publications an author has on a topic, the larger the bubble size. Another type of analysis in the proposed SMS is investigating the number of publications in terms of geographical distribution.

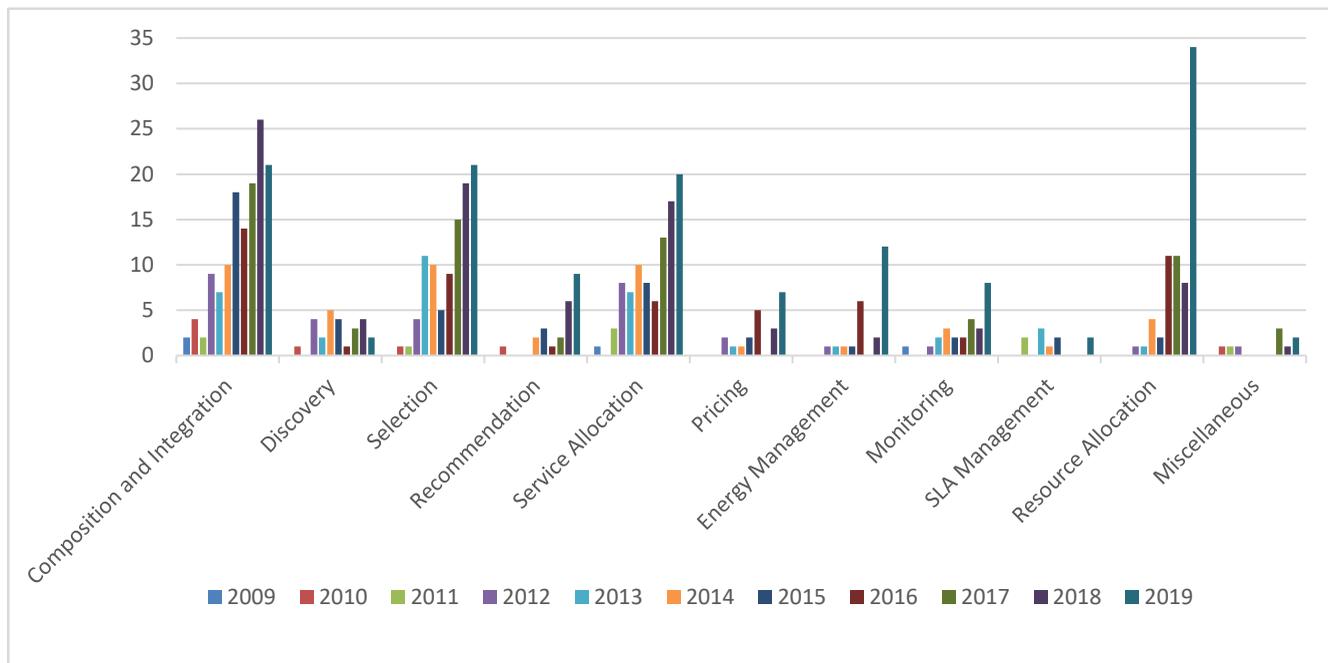

Fig. 3. Published papers per topic per year

Table 7. Most important journals in the field of cloud brokers

| journal title | number of studies |
|---|---|
| future generation computer systems | 80 |
| ieee transactions on services computing | 29 |
| ieee transactions on cloud computing | 29 |
| ieee transactions on parallel and distributed systems | 21 |
| journal of network and computer applications | 18 |
| the journal of supercomputing | 15 |
| cluster computing | 15 |
| journal of systems and software | 11 |
| journal of grid computing | 10 |
| concurrency and computation: practice and experience | 8 |
| ksii transactions on internet and information systems | 7 |
| acm transactions on internet technology | 6 |
| journal of parallel and distributed computing | 5 |
| international journal of computer integrated manufacturing | 5 |

Table 8. Most important conferences in the field of broker

| Conference Title | Number of Studies |
|---|---|
| International Conference on Service Oriented Computing (ICSOC) | 25 |
| International Conference on Utility and Cloud Computing (UCC) | 19 |
| International Conference on Web Services (ICWS) | 16 |
| International Conference on Cloud Networking (CloudNet) | 13 |
| International Conference on Cloud Engineering | 11 |
| ACM/IFIP/USENIX International Middleware Conference (Middleware) | 9 |
| International Conference on Future Internet of Things and Cloud | 7 |
| International Conference on Distributed Computing Systems (ICDCS) | 6 |
| International Conference on Big Data (BigData) | 6 |
| IEEE/ACM International Symposium on Cluster Computing and the Grid (CCGRID) | 6 |
| International Conference on Computing, Networking and Communication (ICNC) | 5 |

Table 9. Number of included and excluded papers in different phases

| Search Space | Initial Set | | Manual & Snowballing | | Database Search-Test | |
|---|---|---|---|---|---|---|
| | Included | Excluded | Included | Excluded | Included | Excluded |
| Journals | 284 | 378 | 42 | 37 | 25 | 95 |
| Others | 133 | 279 | 37 | 108 | 5 | 13 |
| Total | 417 | 657 | 79 | 137 | 30 | 108 |

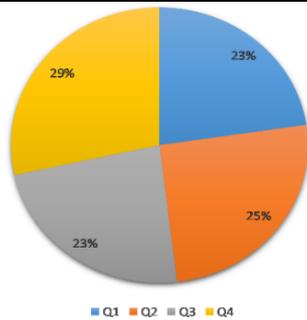

Fig. 4. Included papers per query

This analysis can identify noteworthy institutes and countries that most significantly impact the advancement of the cloud broker field. The data obtained from the extraction phase is used to conduct this analysis. Complete information about the final included study set can be found in $SuppFile_{E1,T5}$. Since a study may have multiple authors with different affiliations, all affiliations are considered here. To answer the geographical distribution RQ for the included papers, author affiliation is considered and the frequency of each country is calculated accordingly.

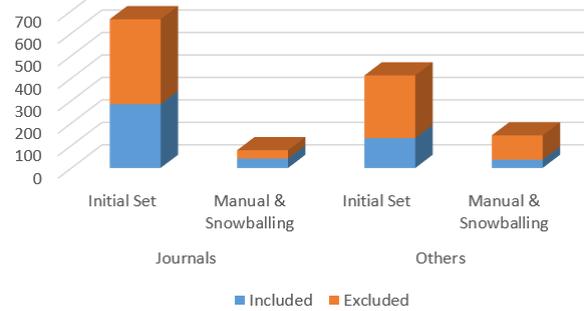

Fig. 5. Total extracted studies per search phase

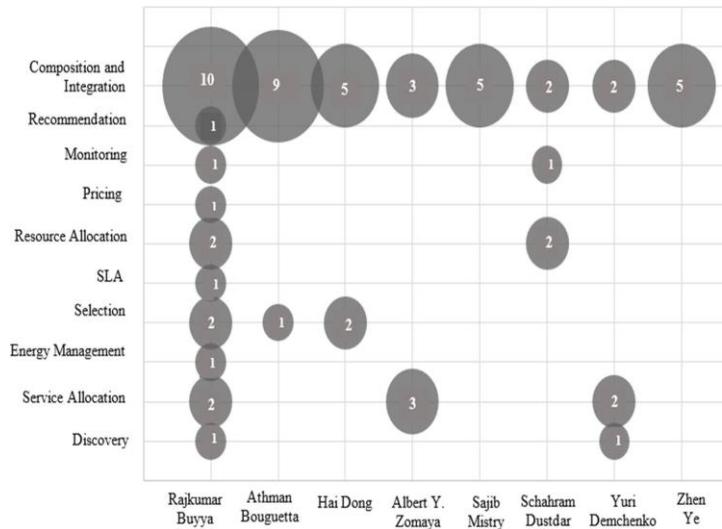

Fig. 6. Active researchers according to the number of publications in the cloud broker's field

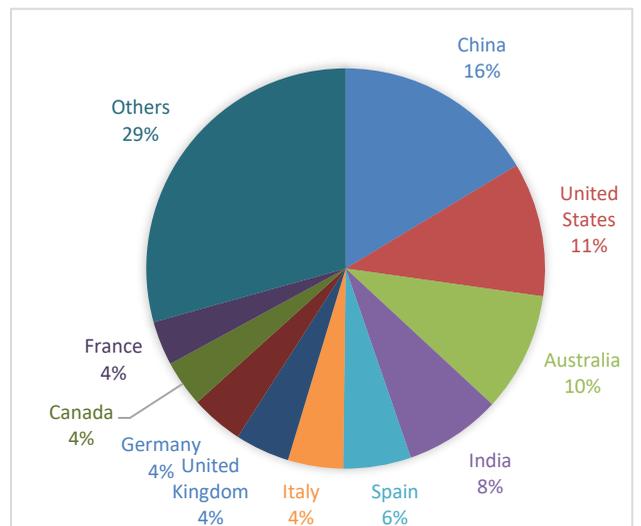

Fig. 7. Geographical distribution of publications

Obviously, in papers with more than one author, the country associated with each author is included in the calculations depending on the affiliation. Fig. 7 depicts the geographical distribution of publications in the field of research. A segment of this figure is named "Others" which aggregates the frequency of all countries having less than 50 publications. According to Fig. 7, China has the largest share (289 publications) of total publications. The United States ranks second. Researchers in these countries have been more active in the field under study, which may be due to the availability of platforms necessary for the cloud and a brokerage infrastructure in research projects and implementation. China, the United States, and Australia conduct more studies because of generously funding researchers and enjoying more technological advancements. In addition, due to the existence of suitable infrastructures in these countries, such as cloud data centers, the field of cloud brokerage has received more attention there. The existence of university research laboratories related to cloud computing, such as the CLOUDS laboratory (managed by Rajkumar Buyya), has also promoted this field.

### 3.3 What are the core research topics in the field of brokering? (RQ3)

Main research topics are found by conducting a topic detection process using clustering keywords. In addition, the present study applies a procedure to reconstruct the research tree and the identification of sub-topics in the cloud broker field. Appendix A, Section 3.4, presents this procedure. The acquired research tree is illustrated in Fig. 1 and comprises 10 main topics as level-one topics. Furthermore, Table A.10 in Appendix A provides some extracted keywords that play a primary role in identifying level-one topics. A comprehensive list of keywords for each level-one topic is found in $SuppFile_{E3}$. The following section describes each level-one topic in detail.

1) Service discovery: With this technology, detecting cloud services and offering appropriate provider resources are automatically done on the internet.
2) Service composition and integration: Service composition and integration concern value-added services and satisfy the demands of users. By considering the type of user demand, a cloud broker gathers all the essential services. Usually, it is possible that all services are offered by just one provider or sometimes it is necessary to combine the different services of various providers. Therefore, a service composition procedure begins with the request of a complex service from the user and then a cloud broker finds and combines all appropriate services according to the quality of services (QoS).
3) Service allocation: In general, the processes involved in providing services intended by the user and/or scheduling tasks on virtual machines fall into this category. A cloud broker can help to optimize the allocation of tasks on virtual machines by providing scheduling services. All of the above concepts are also common in inter-cloud environments.
4) Energy management: Energy management for a cloud provider reduces energy consumption and produces less heat and carbon footprints. The cloud broker can assist in the optimization of energy management by providing appropriate suggestions for running virtual machines on physical machines owned by one or more providers.
5) Service selection: Optimal cloud deployment requires an effective selection strategy that operates according to QoS metrics, such as cost, reliability, and security, and also offers the most appropriate cloud services for end customers.
6) SLA management: Service-level agreement management is one of the challenges in cloud applications. With the advent of sophisticated applications that sometimes lead to the provision of services by several cloud providers, it is crucial for a third-party cloud broker to coordinate between service-level contracts and inter-cloud negotiations to provide the desired QoS.
7) Resource allocation: Resource management is one of the most important issues for a cloud provider. This role includes managing virtual resources on the physical resources of cloud data centers (providing the physical resource and allocating it to the virtual resource) as well as managing other resources, such as storage space and network resources. The cloud broker manages cloud resources as a third party or as a part of the cloud provider. Cloud resource management in a multi-cloud environment can also be performed by the mediation of a broker.
8) Pricing: This concept includes marketing-related mechanisms and the pricing of resources and services of one or more cloud providers. The broker can be involved in processes related to the cloud services market, such as holding auctions, defining new service buying and selling models, and offering profits to providers.
9) Monitoring: Cloud monitoring is a wide term for monitoring diverse aspects of services, from VM performance to the very complex monitoring of cloud services. It should be noted that monitoring systems are needed to track the performance of physical and virtual resources and to run cloud applications.
10) Recommendation: The cloud broker can detect and suggest appropriate services according to the quality of the user's desired service. This helps users choose the right services by making the right offers. A service can also include data management mechanisms.

### 3.4 Which broker topics have the least/most corresponding attention and what is the publication trend and distribution for each topic? (RQ4)

To answer the first part of RQ4, the percentage of publications per topic is computed, as shown in Fig. 8. This figure demonstrates the quota of each level-one topic in all publications in the cloud brokerage field. As seen, the composition and integration topics attract the most attention. The topics of selection and service allocation are the second and third most important research topics in the cloud broker field, respectively. The degree of researcher reception to research topics on discovery, recommendation, energy management, and monitoring are almost equal. Among all the extracted level-one topics, SLA management and pricing show the lowest publication rate.

Confirming the information presented in Fig.8, Fig.9 depicts the evolution of each topic over time showing that, among all the extracted level-one topics, composition and integration and selection are the most popular. In cloud computing, some services, such as service allocation, service composition, and service selection, are considered as the most basic duties of a cloud broker. Accordingly, these topics attract more attention in research. Furthermore, there are some other important topics, such as discovery and monitoring services, which require complex algorithms for implementation. Hence, the research works conducted on these topics are fewer than those of the first group. The third group consists of topics, such as pricing and recommendation. These services are higher level services and are expected to be of more interest to broker system researchers in the future. Furthermore, energy management is a critical area, but has so far received comparatively less attention from researchers; new work in this area is expected. Fig.9 and Fig.10 present the evolution of research studies from both perspectives, i.e., client-centric and provider-centric.

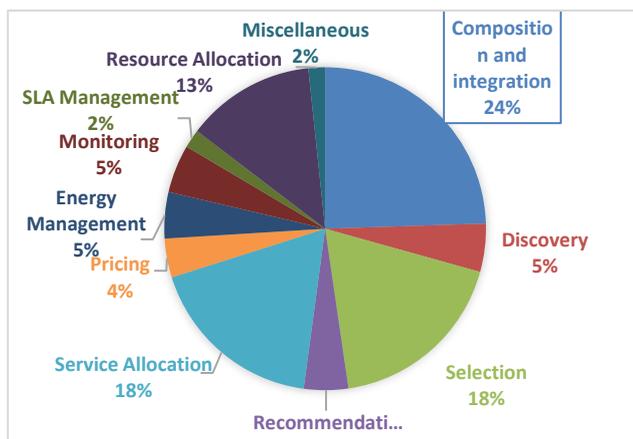

Fig. 8. Percentage of publications per topic

### 3.5 Which forms of empirical evaluation have been used? What are the tools available to support field approaches? Which techniques are used more in the field? (RQ5)

Additional valuable information extracted from the proposed SMS regards investigating the popularity of evaluation methods in cloud brokerage. The empirical evaluation features testbed, practical, and simulation. Testbed is an implementation of a real cloud on a smaller scale by utilizing cloud management software, such as OpenStack. Also referred to in articles as implementation, Practical is a real implementation in a commercial cloud. Simulation is a cloud simulator, such as CloudSim, or is a self-development, in which a problem is solved by existing programming languages, such as Java and Python. Fig.11 presents the types of evaluation methods employed in the studies reviewed.

Furthermore, since all the techniques used in the included studies are considered, answering this RQ will familiarize researchers interested in cloud brokerage with this field's methods for satisfying user demands.

Most studies propose an architecture or framework for search studies. Ranking next are heuristic and meta-heuristic algorithms. Fig.12 presents the techniques used by brokers for solving current issues in cloud environments

### 3.6 What is the relation between topics and broker roles in the NIST category? Which NIST roles have the least/most corresponding attention? (RQ6)

As seen in Fig.13, out of 496 included studies, 466 studies are related to NIST categories, the majority of which (70%) have paid special attention to the role of intermediation. In comparison to intermediation, as the simplest and currently the most operational type of broker, aggregation and arbitrage are relatively new concepts. In addition, intermediation is a primary type of broker and performs the most basic broker task, namely to find a service. In the future, however, with the advancement of broker capabilities, the concepts of arbitrage and aggregation are expected to be further explored as broker duties. The addition of such features will enable brokers to combine different services to meet complex requests. Accordingly, an analysis of the 2009-2019 studies reveals that 79 and 55 studies identify arbitrage and aggregation as new broker capabilities, respectively.

As described in Section 2, by examining the included studies, the proposed SMS extracts a set of important topics in the broker field (level-one topics in Fig.1). The purpose of answering this research question (RQ6) is to investigate the relationship between the topics extracted and the role of broker, which is classified by NIST into three general categories: aggregation, arbitrage, and intermediation. Reflecting this, Fig.14 shows that the number of included papers considering composition and integration also address, among other topics, the roles of aggregation and arbitrage. However, most researchers have focused on intermediation because of its more prominent role.

### 3.7 In which environment and service layer is the broker mostly considered? (RQ7)

In the systematic review of the broker field, one of the vital analyses is to consider the level at which the broker plays a role. Based on the present study's analyses, the levels of cloud services are divided into five categories: SaaS, PaaS, IaaS (IaaS and CaaS), XaaS (all: anything as a service), and Other (special services, e.g., NaaS (Network as a Service) and DaaS (Desktop as a Service)). Since most commercial cloud services are in the IaaS layers, the result illustrated in Fig.15 is predictable and also acceptable. It should be noted that, if a type of cloud service in the included studies has not been explicitly indicated, then the current work has considered that service as XaaS, which generally covers any type of cloud service.

### 3.8 What is broker control orientation? (RQ8)

In this RQ, the present study analyzes the broker from the viewpoint of a centralized or distributed orientation. In studies, broker implementation is divided into two categories: distributed and centralized. In the centralized model, a single broker entity is responsible for managing the broker's tasks, while, in the distributed model, a number of brokers perform brokerage tasks in coordination with each other. It is clear that the centralized model is easier to implement, and, since all information is stored in the centralized broker, the decision making is also easier. However, the disadvantage of this model is the unwillingness of brokerage process participants to provide their information to the broker or the difficulty for a centralized broker to have all the information Fig.16 presents the types of cloud brokers. As seen in Fig.16, a large number of included studies have used centralized algorithms to present and demonstrate cloud broker capabilities. Accurate and instantaneous data collection is difficult for a centralized broker and some participants in the system, such as providers in a federation, may be reluctant to share their information with a centralized broker. In this case, the distributed model is essential since each provider has its own broker and provider privacy is maintained. However, each provider must estimate the information of other system members and make decisions based on its own local data and just a small amount of the other providers' information.

## 4 DISCUSSION ON THE RESULTS

This section compares the proposed SMS with the review systems of other articles in the field and discusses the degree of completeness of each. The following sections will further review and analyze the results of Section 3, i.e., evaluation methods utilized by cloud brokers, cloud broker applications, and approaches for broker development.

### 4.1 What is the difference between our SMS and other reviews? In comparison to previous studies, to what extent our SMS covers the main topics in this area?

The main objective of answering this RQ is to provide a comparison between the proposed SMS and other related reviews. As mentioned previously in Section 1, the present study selects three review papers [38],[28],[23] from Table 1, which are the most similar to its own work, for the purpose of conducting a comprehensive comparison. Table 10 illustrates the comparisons made. As seen, the current review investigates a large set of search spaces and selects high-quality related papers among a large pool of extracted papers. In addition, a comprehensive search strategy is presented which can find a majority of related search spaces and studies and then include/exclude papers in consideration of some exclusion criteria presented in the Appendix A briefly described in Section 2. In the current study, all of the investigated secondary works are placed in the range of 2012 until 2019. However, most lack a systematic methodology to cover all relevant studies in the cloud broker field. Among the 24 secondary articles, three are selected for deeper investigation because of their closer similarity to the desired fields and greater comprehensiveness. The selected surveys are Fowley et al. [28], Chauhan et al. [23], and Elhabbash et al. [38]. As can be deduced from Table 10, the proposed SMS (last row) investigates a large set of related papers for its comparison to other review studies. . Most of these review studies do not offer a systematic process to review articles on cloud brokerage and each has only investigated a small set of papers in the desired field. In contrast, the proposed search strategy introduces a comprehensive process for finding a complete set of related papers and for selecting a set of high-quality papers in the cloud broker field. The present search strategy is a three-tier process consisting of a manual search, snowballing search, and database search. In the manual search phase, each venue in the search space list (acquired from investigating references of existing reviews in Table 1) is manually searched using a set of constructed queries. Each query consists of a set of keywords found in $SuppFile_{W3,T3}$.

Table 10 presents another comparison between the proposed SMS and other reviews in regard to the topics presented.

### 4.2 The trends and demographics and active search spaces in cloud broker research (RQ1, RQ2, RQ3, and RQ4)

Observing the demographics of cloud broker research uncovers the importance of the broker in studies conducted on cloud environments. The upward growth of the bar charts in Fig.2 (RQ1) indicates that the cloud broker has been widely accepted as one of the promising solutions for cloud environments.

### 4.3 The trends and demographics and active search spaces in cloud broker research (RQ1, RQ2, RQ3, and RQ4)

Observing the demographics of cloud broker research uncovers the importance of the broker in studies conducted on cloud environments. The upward growth of the bar charts in Fig.2 (RQ1) indicates that the cloud broker has been widely accepted as one of the promising solutions for cloud environments. The present study has identified key researchers and research venues in the cloud broker field that can serve as guidelines for those intending to research this area. A complete list of research venues is available in $SuppFile_{W3,T4\ and\ T6}$. It should be noted that knowing the geographical distribution of publications (shown in Fig.7) can be helpful in finding the key geographical locations of cloud broker researchers. There is a direct relationship between the volume of research conducted in cloud brokerage obtained from Fig.3, the topics of service composition, service selection, and resource allocation show the highest number of research studies and publications until 2019 when compared to other broker topics. With the spread of startups, the need for cloud resources is growing. The usage of a resource allocation strategy by providers will facilitate more effective utilization of cloud resources and optimize the revenue generated from these.

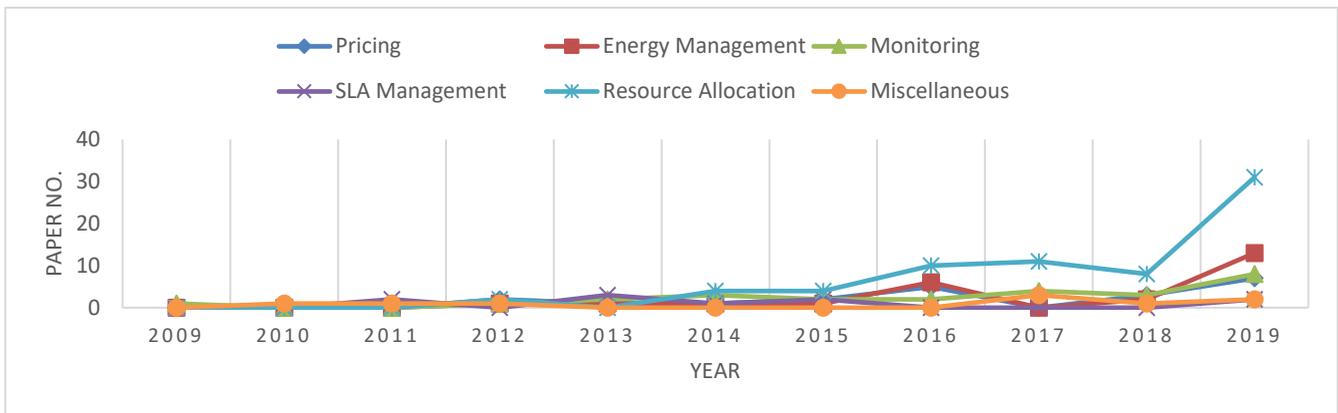

Fig. 9. Evolution of the publication of research studies in client-centric topics

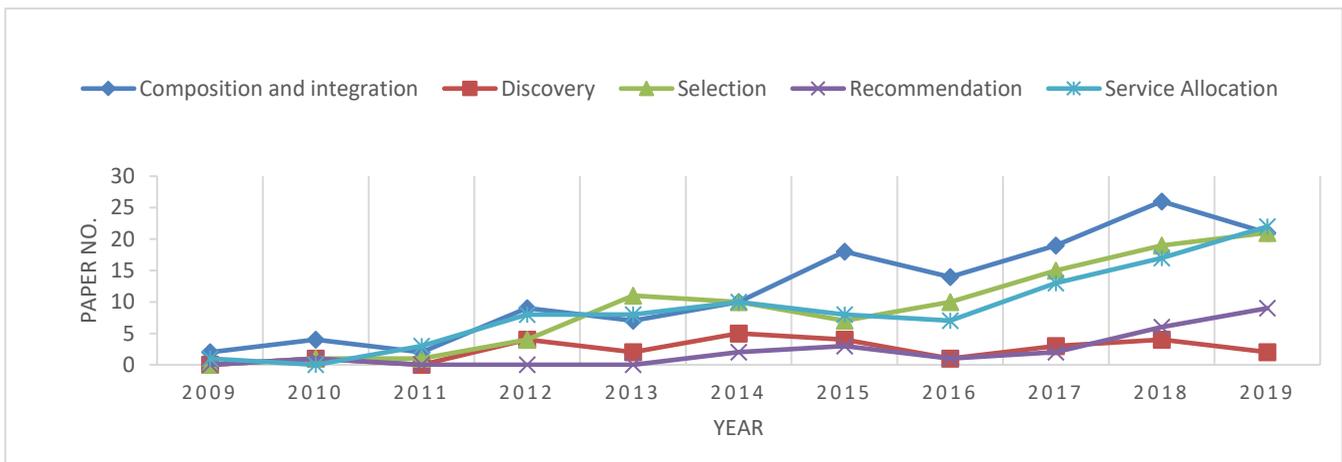

Fig. 10. Evolution of the publication of research studies in provider-centric topics

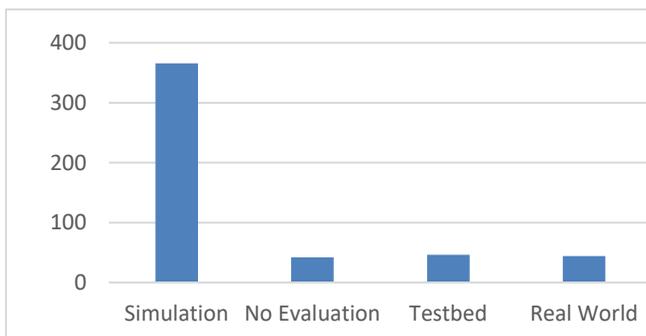

Fig. 11. Empirical evaluation

Fig.3 demonstrates that a significant amount of research has been conducted with the aim of introducing effective solutions for the topics of composition and integration, selection, and service allocation.

To meet the needs of customers in cloud environments, the allocation of resources is provided in a web platform and is kept elastic and virtual. With the expansion of cloud environments, many commercial companies compete to sell their cloud resources and to attract customers. Among this large volume of cloud services, it has become a great challenge to find the most suitable service that is tailored to the needs of users. Cloud brokers have taken on this challenge as one of their most critical tasks and so have found a special niche in the cloud environment. Considering the acquired results presented in Fig.9 and Fig.10, 15% of the included studies have suggested solutions for selecting the appropriate service when taking into account the preferences of cloud customers. According to Fig.8, 101 studies have been conducted on the selection of cloud services, in which the defined objectives comply with the NIST standards on brokers. On the other hand, depending on customer requirements, cloud providers offer different types of services which often need to be combined. Therefore, the use of service composition is growing as a popular technology to composite and integrate distributed and heterogeneous services. The most important advantages of applying the service composition technique are reducing costs and time as well as improving performance. The composition of cloud services did not take place in the early days of cloud brokerage (in 2009).

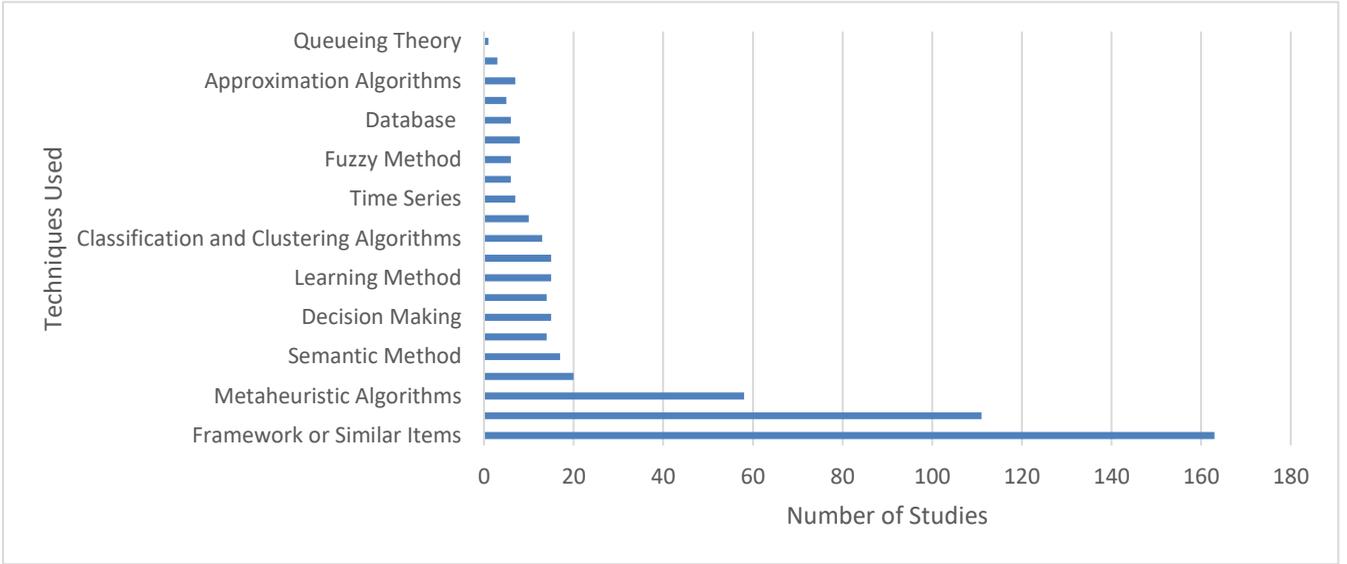

Fig. 12. Techniques used to implement brokers in the included studies

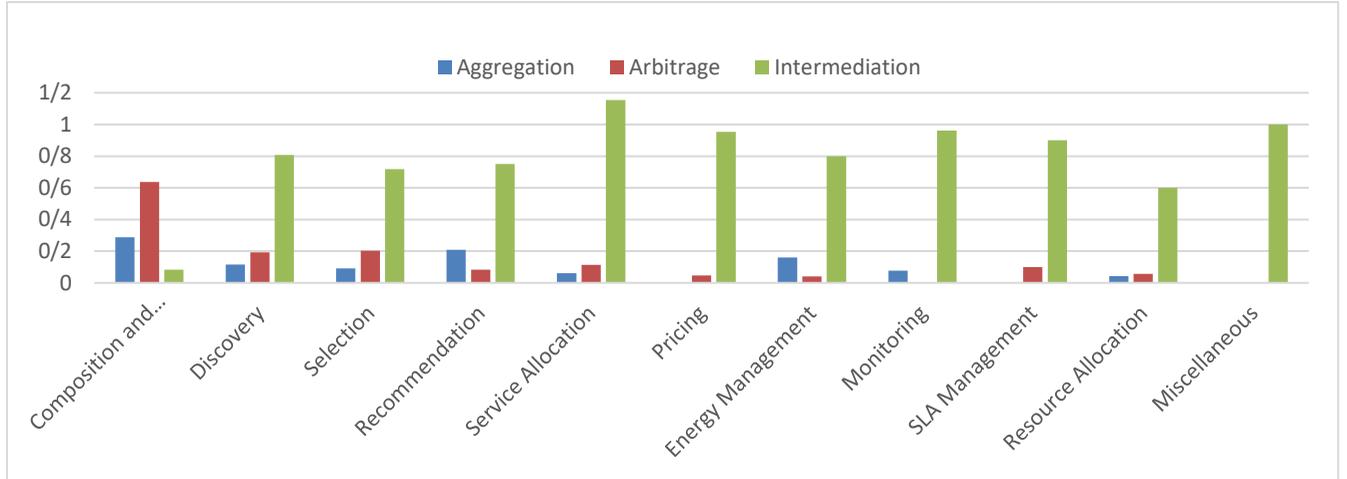

Fig. 13. Relation between extracted topics and NIST roles

Table 10. Comparison between the proposed sms and other reviews

| Ref. | Review Type | #Studies | #Search Spaces | #Time Interval | #Considered Topics/Sub-Topics |
|---|---|---|---|---|---|
| **[33]** | Survey | NM[1] | 17[2] | 2012 to 2015 | Aggregation Customization Intermediation Integration Arbitrage |
| **[38]** | Survey | 47[2] | 37[2] | 2013 to 2018 | Pricing Multi-Criteria Quality of Services Optimization Trust |
| **[39]** | Survey | 34 | 30 | 2010 to 2017 | Decision Support Resource Monitoring Policy Enforcement SLA Negotiation Application Deployment Migration API Abstraction VM Interoperability |
| **Proposed SMS** | SMS | 496 | 171 | 2006 to 2019 | Service Composition & Integration Service Discovery Service Selection Energy Management SLA Management Resource Allocation Pricing Recommendation Service Allocation Monitoring |

[1] The value is Not Mentioned (NM) in the paper.
[2] The value is not mentioned directly in the paper but can be extracted manually.

Table 11. Comparison between the proposed sms and related reviews regarding extracted level-one topics

| References | [39] | [16] | [7] | [28] | [17] | [37] | [40] | Our SMS |
|---|---|---|---|---|---|---|---|---|
| opics | | √ | | √ | √ | | | √ |
| Composition & Integration | | | | | √ | | √ | √ |
| Discovery | √ | | √ | √ | | | √ | √ |
| Service Allocation | | | | | | | | √ |
| Energy Management | | √ | √ | | | | √ | √ |
| Selection | √ | | √ | | | | √ | √ |
| Resource Allocation | √ | √ | | | | | √ | √ |
| Pricing | √ | √ | | | | | √ | √ |
| Monitoring | | | √ | | | | | √ |
| Recommendation | √ | √ | | | | √ | √ | √ |
| SLA Management | | | | | | | | |

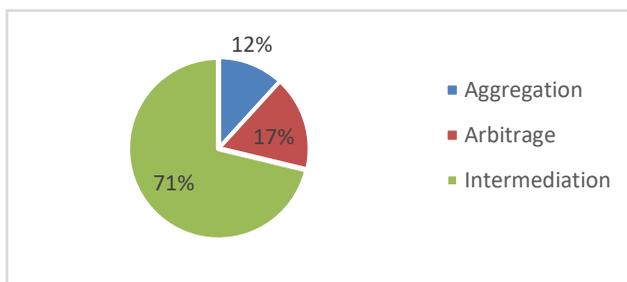

Fig. 14. Frequency of NIST roles in the included papers.

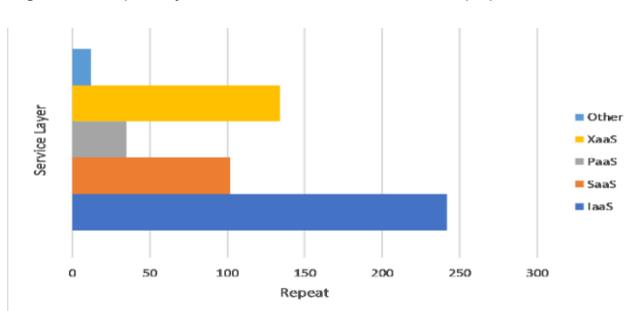

Fig. 15. Types of service layers in cloud computing

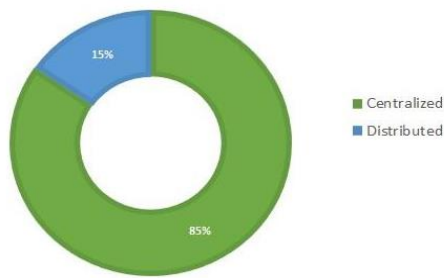

Fig. 16. Types of brokers

In fact, it was not until 2012, with the advent of inter-cloud environments, that the topic of service composition started gaining much importance due to the complex demands of customers, expansion of cloud environments, and rising competition among providers in offering better services.

Fig.3 demonstrates that, from 2009 to 2019, the trend of publishing studies on composition and integration was in an ascending direction. The results of Fig.8 reveal that, among the main broker tasks (level-one topics extracted from the research tree), researchers focused on providing effective solutions for combining and integrating services (the topics of composition and integration respectively). Included in the topic of composition and integration were 133 published studies whose objectives comply with the composition defined by NIST (Fig.13 in RQ6). More precisely, out of 133 studies on the topic of composition and integration, 63% are related to the role of arbitration, 28% to integration, and less than 10% to intermediation. As a guideline for researchers interested in brokerage, it can then be said that, according to the analysis of the included studies, composition and integration, resource allocation, and service selection are the more outstanding and active topics in this field. Among the three main roles of aggregation, arbitration, and intermediation introduced by NIST, the largest number of research studies published from 2009 to 2019 were on intermediation (RQ6). Because intermediation is the simplest and most operational task of the broker, integration and arbitration are new phenomena that have been welcomed and considered by many researchers since 2013 (RQ6).

### 4.4 Research efforts towards cloud broker evaluation (RQ5)

The results of RQ5 reveals that the majority of studies (71%) have applied simulations to evaluate their methods, and a small number of studies have employed testbed (10%) and real-world methods (10%). It should be noted that 10% of the research studies have not mentioned the type of evaluation used. Real-world methods usually offer more benefits because of being more accurate and less exposed to the bias and manipulation of parameters. Real-world methods can also serve as a useful indicator of what actually exists in cloud environments. However, due to the dynamic nature of execution time conditions, the application of real-world methods to test and evaluate proposed solutions is difficult and costly. These challenges have led to the usage of simulation methods, such as CloudSim, for evaluating research approaches. On the other hand, the application of simulation methods has caused the actual conditions at the execution time to be ignored and created a gap between **"what has been evaluated"** and **"what actually exists."**

As mentioned earlier, due to the dynamic conditions of the execution time, most researchers have employed simulation methods to evaluate their work. However, researchers must pay attention to the objectivity of the evaluation method used. For example, researchers utilizing simulation methods to evaluate their solutions should prove the objectivity and quality of their proposed solutions to industry experts. Fig.11 illustrates a variety of techniques for designing a cloud broker, including metaheuristic algorithms, frameworks and similar items, and semantic and fuzzy methods. Evaluating such techniques in the real-world may generate different results than when evaluating with simulation methods. Of the work observed, Fig.11 reports that 31% of the studies have used frameworks and similar items to construct a cloud broker. Ranking second in this figure, at 21%, are studies utilizing metaheuristic algorithms to perform broker tasks.

### 4.5 Research efforts towards cloud broker usage (RQ3, RQ5, and RQ7)

The most common task of a broker is to meet the needs of both cloud providers and customers. To achieve this objective, researchers should be aware of the current conditions of cloud environments in order to construct a perfect cloud broker. To apply the techniques introduced in research studies, it is necessary to have special conditions. For example, with the expansion of cloud environments and the rise in customer expectations, often a cloud service alone is unable to meet the needs of customers. Hence, the use of techniques introduced in the construction of brokers is essential for selecting cloud services and combining them. The analysis of RQ4 reveals that, in 31% of the research studies conducted, the topic of composition and integration is considered as one of the most important tasks of the broker. Also, 19% of the included studies focus on the selection of cloud services through the broker. According to the analysis of the included studies, the primary responsibilities of the broker are divided into ten important and primary tasks, each of which is described in RQ3. Interpretations of RQ5 reveal that a significant portion of the research is focused on how to make and apply brokers in large-scale cloud environments. On the other hand, the results of RQ8 indicate that, in large-scale cloud environments with high complexity, the use of distributed brokers is generally more common than that of centralized brokers. Since the centralized broker is the most basic type of broker and its implementation is easy, much research has been conducted in this field, as shown in Fig.16. It should be noted that, since the centralized broker itself holds all the required information, it has a simple function which has attracted the attention of many researchers. In contrast, a distributed broker does not need to have all the information to make a decision and can perform its job independently of other brokers. feature of distributed brokers is interesting and should be explored further in the future

### 4.6 Research efforts towards cloud broker development (RQ7, RQ8)

Since 2006, with the advent of cloud computing, IaaS services have been the first to be provided for customers on the Internet. Over time, various levels of cloud service have emerged, which the present study has divided into five categories in RQ7: SaaS, PaaS, IaaS (IaaS and CaaS), XaaS (all: anything as a service), and Other (special services, e.g., NaaS (Network as a Service) and DaaS (Desktop as a Service)). The analysis of the included studies concluded that, out of the five cloud service categories, most researchers are inclined to provide brokers in the IaaS layer due to the popularity or prevalence of their services over those of the other service layers. According to the results obtained from RQ7, Fig.15 demonstrates that 46% of the studies propose the cloud broker in the IaaS layer. Approximately 20% of the studies introduce the SaaS layer cloud broker, while the rest of the studies (34%) consider the broker in other levels.

Another important aspect that should be considered is the centralization or distribution of brokers in cloud environments (RQ8). Generally, the most basic and simplest broker is a third-party broker that communicates between cloud customers and service providers. The centralized broker is the simplest broker to be introduced and used in cloud environments. Section 4.4 and Section 3.8 compare the two types of brokers in detail.

## 5 IMPLICATIONS OF THE FINDINGS

The present study has carried out a systematic review on cloud broker research for the purpose of guiding researchers, stakeholders, and educators interested in this field. Due to the wide range of search spaces under review, there are appropriate and worthy implications for different research audiences from the results and discussions presented in Section 3 and Section 4. Each of the results presented in the previous sections can significantly assist various audiences in this research field. In this section, the implications of the proposed SMS are presented for researchers, stakeholders, and educators.

### 5.1 Implications for researchers

- There are relatively large differences among the rates of studies conducted in different countries (RQ2). China (289 studies), the United States (191 studies), and Australia (170 studies) are among the most active countries in this field of research. The present study has deduced that the high volume of research in these countries is due to the advancement of technology in their industries in comparison to that of other countries as well as the existence of a strong relationship between academic centers and industry. With the advancement of a country's technology and industry, it is critical that academic research is conducted to achieve efficient methods and to meet the various needs of industry. Therefore, the degree to which a national industry progresses in its usage of cloud technologies to meet user requirements will directly impact the level of acceptance and motivation of researchers in that country to conduct studies in this field. On the other hand, to introduce efficient methods for the cloud broker, researchers must employ a suitable platform for evaluating these approaches before entering the realm of industry. Consequently, a vital step towards this aim is allocating budgets that provide the appropriate infrastructure to evaluate academic research aimed to advance industry goals. As RQ5 reveals, those countries with a suitable and powerful infrastructure, such as cloud data centers, can better support and implement research work in the industry. Furthermore, the existence of research laboratories dedicated to cloud computing, such as the CLOUDS laboratory managed by Rajkumar Buyya, has also been influential.

- Combining cloud services to meet the complex needs of users is possible in both single-cloud and multi-cloud environments. However, in consideration of the extent of inter-cloud environments, environmental conditions at the time of the broker's implementation in these environments are more variable and unpredictable than in single-cloud environments. Therefore, when a broker is an orchestration, in addition to selecting the service, it should appropriately combine services and increase the resistance to failure.
- Due to the widespread use of cloud services by cloud customers and startups, as well as the broker's requisiteness to combine services to provide better services, it is imperative that failures be managed during the broker's execution. This is critical in the process of combining services, as the failure to run only one service will cause the process of running the broker to fail. Therefore, as a guideline for researchers and audiences interested in brokerage, assiduity is vital to the mechanisms of management and the detection of failures during the implementation of web service composition.

### 5.2 Implications for stakeholders and practitioners

- The results of RQ1 illustrate that brokers have been on the cloud computing scene since 2009 and will soon figure as one of the main components of cloud computing negotiation and business-to-market cloud services. In regard to the commonly utilized solutions of recent decades, stakeholders should take advantage of this acquired experience to improve cloud-based services.
- Current research on the development and use of cloud brokers is theoretical and academic. In industry, it is not common practice to apply solutions proposed in academic studies. Cloud practitioners and stakeholders must play a key role in improving the current technology in cloud computing. To achieve this, practitioners in the area of cloud computing, especially in the broker field, should participate in the most respected academic conferences and workshops on this topic (RQ1). The presentation of their perspectives and preferences on current and future research approaches can significantly promote the advancement of cloud computing in the industry. The presence of these stakeholders and practitioners at such gatherings will greatly influence the orientation of algorithms presented by researchers so as to adapt to the dynamic and real conditions of cloud environments. The experience of practitioners can significantly affect the method of constructing and classifying brokers (RQ2). In turn, practitioners can profit by collaborating with researchers in academia.
- Presently, most solutions offered in broker research have adequate quality for use in real-world environments (RQ1, RQ6). However, there is a lack of empirical evidence from industry. Industry experts (practitioners) and stakeholders should collaborate with researchers to widen the horizon of academia in industry-friendly metrics.

### 5.3 Implications for educators and teachers

With over 10 years of experience, cloud brokers are becoming one of the most promising solutions for trading in complex cloud environments (RQ1). These constructed brokers are the result of studies by some of the most respected researchers and pioneers in the field of cloud brokers. Of course, an advantage of utilizing academic experience in constructing an industry-friendly broker is the transfer of knowledge to novice developers. As a guideline, the results of the proposed SMS suggest that instructors of courses on, for example, cloud computing and distributed systems, should include cloud brokers in their syllabus as an important component of cloud environments. Extensive research conducted in the broker field can be utilized as a training resource for those teaching cloud computing. The analysis of the included studies in Sections 4 and 5 points out that educators can inform students about a variety of unpredictable conditions and the occurrence of failures and faults that may happen during broker implementation.

## 6 CONCLUSION AND FUTURE WORK

The present study conducts a systematic mapping study on cloud brokers that pertains to the time frame of 2009 to the end of 2019. A total of 1,298 related studies from search spaces are extracted and then 496 studies are selected based on the quality criteria established in the search strategy. The references of the 496 included papers are in Appendix B. An important part of the proposed SMS is the presentation of a powerful research methodology. The introduced SMS contains a comprehensive three-tier search strategy consisting of a manual search strategy, backward snowballing, and database search of reputable scientific libraries. The accuracy of the search methodology has been analyzed in terms of extracting related studies and collecting comprehensive and complete information in a supplementary file and also, the detailed explanation of the reviewing process is inserted in Appendix A. The evaluation results of the search methodology report that more than 80% of the studies published in the broker field can be found. Also provided is a comprehensive supplementary document containing complete details of the information extracted and reviewed in each phase of the current study's systematic review.

A set of eight research questions are determined and responded to during the proposed SMS. The first three research questions (RQ1, RQ2, and RQ3) address reviewing the amount of research conducted in the broker field from 2009 to 2019, extracting the most important topics and tasks of the cloud broker, and introducing the most important researchers and pioneers in the cloud broker field. In addition, by answering RQ1, RQ2, and RQ3, broker audiences will become acquainted with countries active in the broker field. Two other questions, RQ4 and RQ5, investigate the amount of research conducted in each of the important broker topics, the rate of research growth in each topic over time, the types of evaluation methods used in research studies, and, finally, the techniques applied to construct cloud brokers. The present study also compares the proposed SMS against several new and valid related reviews and examines the depth of the methodology used, the number of studies investigated, and the various aspects considered in each conducted review. In the last three questions (RQ6, RQ7, and RQ8), the tasks and topics extracted from the included studies are examined and analyzed from the perspective of NIST definitions. In RQ7, the current work identifies the different layers of service for the cloud broker and determines the number of included studies considered in each layer. In RQ8, two important aspects of cloud brokers are examined, namely centralized and distributed, while broker studies are analyzed according to two defined aspects.

The proposed SMS demonstrates that the cloud broker field is active and growing in various geographical locations and that the development of cloud brokers needs to occur in conjunction with the latest research achievements. Research is increasingly employing brokers to develop interactions between customers and cloud service providers. Systematic mapping studies, such as the proposed SMS, can be utilized as a basis for a more specific review of systematic literature. In future work, each of the top-level topics extracted from the introduced research tree can be further explored to answer more specific research questions.

## Biography

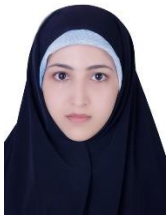
**Hoda Taheri** received the B.S. and M.S. degrees in computer engineering, with concentration in parallel and distributed systems, and clustering wireless sensor networks, respectively. she is currently Ph.D. candidate of computer engineering in Ferdowsi University of Mashhad (FUM). Her research interests focus on resource management and scheduling in distributed systems, especially in clouds.

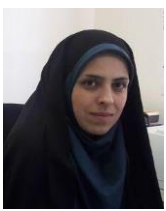
**Faeze Ramezani** received her B.Eng in Computer Engineering from Ferdowsi University of Mashhad (FUM) in 2007, MSc degrees in computer engineering from the University of Isfahan, in 2011 and she is currently Ph.D. Candidate of Computer Engineering – Software in FUM. Her research interests are resource management in cloud, game theory and mechanism design.

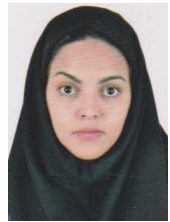
**Neda Mohammadi** is a PhD Student at the Software Quality Laboratory (SQL) Lab, in the Department of Computer Science and Software Engineering at Ferdowsi University of Mashhad, Iran. Her research interests include microservice-based systems with an emphasis on microservice patterns, empirical software engineering and software quality, recommendation systems, and web service composition with an emphasis on fault tolerance. You can find more information at http://n.mohammadi.student.um.ac.ir/.

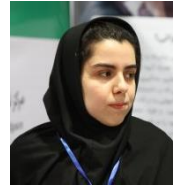
**Parisa Khoshdel** received her BSc degree in computer engineering-softwate from Kosar University, Iran and her MSc degree in Computer engineering from Ferdowsi University, Iran in 2021. Her current research intrest lies in cloud computing and game theory.

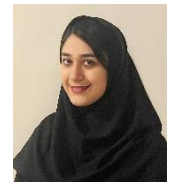
**Bahareh Taghavi** received the B.Eng. degree in Computer Engineering from Golestan University, and the M.Eng. degrees in computer engineering from the Ferdowsi University of Mashhad (FUM), Iran in 2021. Her current research interests include cloud computing and workflow scheduling.

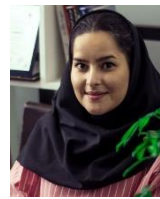
**Neda Khorasani** received her B.Eng and M.Eng degrees in Software Engineering from Ferdowsi University of Mashhad, in 2016 and 2019, respectively. She has high-level experiences and interests in cloud computing, resource management, scheduling, and game theory.

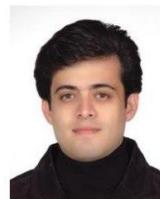
**Saeid Abrishami** received the MSc and PhD degrees in computer engineering from Ferdowsi University of Mashhad, in 1999 and 2011, respectively. Since 2003, he has been with the Computer Engineering Department, Ferdowsi University of Mashhad, where he is currently an assistant professor. During the spring and summer of 2009 he was a visiting researcher at the Delft University of Technology. His research interests focus on resource management and scheduling in the distributed systems, especially in Grids and Clouds.

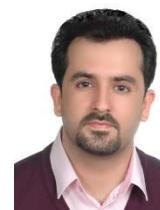
**Abbas Rasoolzadegan** received his B.Sc. degree in Software Engineering from Aeronautical University,Tehran, Iran, in 2004. He also received M.Sc. and Ph.D. degrees in Software Engineering from Amirkabir University of Technology, Tehran, Iran, in 2007 and 2013, respectively. He is currently an Associate Professor with the Computer Engineering Department of Ferdowsi University, Mashhad, Iran. His main research focus is on software quality engineering, especially in terms of design patterns and refactoring. For more details, visit the SQLab homepage at http://sqlab.um.ac.ir. appears here